\documentclass[
  aps,prl,twocolumn,superscriptaddress,nofootinbib,floatfix
]{revtex4-2}

\usepackage[T1]{fontenc}
\usepackage[utf8]{inputenc}
\usepackage{amsmath,amssymb,mathrsfs}
\usepackage{graphicx}
\usepackage{xcolor}
\usepackage{microtype}
\usepackage{needspace}
\usepackage[colorlinks=true,allcolors=blue]{hyperref}
\newcommand{\scri}{\mathscr I^+}

\newcommand{\GRA}{\texttt{GR-Athena++}}

\begin{document}

\title{Hyperboloidal Evolution of the Einstein Equations at Null Infinity for Gravitational-Wave Extraction and Matching}
\author{An{\i}l Zengino\u{g}lu}
\affiliation{Institute for Physical Science and Technology,
University of Maryland, College Park, Maryland 20742, USA}
\author{Andrea N\"utzi}
\affiliation{Department of Mathematics, Stockholm University,
SE-106 91 Stockholm, Sweden}
\author{Sebastiano Bernuzzi}
\affiliation{Theoretisch-Physikalisches Institut,
Friedrich-Schiller-Universit\"at Jena, 07743 Jena, Germany}

\begin{abstract}
We present a new scri-fixing approach for the (3+1) Einstein equations on spacelike slices extending to null infinity. The method enables accurate computation of gravitational radiation, including asymptotic charges and nonlinear memory. We demonstrate Cauchy-hyperboloidal waveform extraction for a binary-black-hole merger, finding subpercent agreement with characteristic extraction. We present the first Cauchy-hyperboloidal matching computation for a perturbed black hole, resolving quasinormal ringing and late-time tails, and enabling astrophysical applications.
\end{abstract}

\maketitle

\emph{Introduction.}---%
Gravitational-wave information about strong-field dynamics radiates far beyond the outer boundary of finite-domain simulations. The waveform
and related observables
are naturally defined at the ideal boundary reached by outgoing light: future null infinity, $\scri$~\cite{Penrose:1962ij}. A slice is
called
\emph{hyperboloidal} if it approaches an outgoing light cone at large radius and is spacelike
up to and including the two-dimensional cut where it meets $\scri$. On such a slice, conformal compactification maps the entire wave zone  onto a finite grid, but the vanishing conformal factor introduces formally singular terms in the Einstein equations at $\scri$. Regular formulations of the hyperboloidal initial value problem remove these singularities~\cite{Friedrich1983,Friedrich:1986qfi,Kroon:2016ink}, but maintaining numerical resolution
near $\scri$
during \emph{long-term} evolutions for astrophysical applications has remained a challenge.

In this Letter, we combine a recent regular orthonormal-frame formulation of the Einstein equations~\cite{Nutzi:2025kqc,Nutzi:2023uly} with scri-fixing~\cite{Zenginoglu:2007jw} and hyperboloidal layers~\cite{Zenginoglu:2010cq,Bernuzzi:2011aj,Bernuzzi:2025hhu} to design a practical approach to the hyperboloidal initial value problem
for astrophysical simulations. In this formulation the location of $\scri$ is independent of the solution. Scri-fixing places it at a time-independent grid radius, so the physical domain retains its radial grid points as the radiation decays.

Nonlinear conformal evolutions based on the hyperboloidal framework proposed by Friedrich \cite{Friedrich1983,Friedrich:1986qfi} began in the 1990s~\cite{Hubner:1994pd,Frauendiener:1997ze,Husa:2002kk}. H\"ubner evolved a three-dimensional vacuum spacetime through null and timelike infinity~\cite{Hubner:2000pb}.
Subsequent developments include constrained evolution~\cite{Moncrief:2008ie,Rinne:2009qx}, tetrad formulations~\cite{Bardeen:2011ip}, the dual-frame formulation~\cite{Hilditch:2015qea,Hilditch:2016xzh,Gasperin:2018tvk,Duarte:2022vxn,Peterson:2023bha,Peterson:2024bxk,Peterson:2025csg}, free evolution in spherical symmetry~\cite{Zenginoglu:2007it,Zenginoglu:2008pw,Vano-Vinuales:2014koa,Vano-Vinuales:2017qij,Vano-Vinuales:2023yzs,Vano-Vinuales:2024tat,Alvares:2025jjm,Vano-Vinuales:2025oai}, nonlinear black-hole scattering~\cite{Frauendiener:2021eyv,Frauendiener:2022bkj,Frauendiener:2023ltp,Frauendiener:2025xcj}, and 3D evolution of perturbed Minkowski spacetime with scri-fixing~\cite{Vano-Vinuales:2026izp}.
However, accurate hyperboloidal simulations of the fully nonlinear Einstein equations for astrophysical radiation remained an open problem. Here we demonstrate a practical solution to this problem: we extract the waveform of a binary-black-hole merger at future null infinity and evolve a perturbed single black hole using two-way Cauchy--hyperboloidal matching.

A more common approach to including null infinity in numerical relativity is the characteristic approach \cite{Bishop:1996gt, Winicour:2012znc}.
Characteristic methods already provide binary waveforms and memory at null infinity through Cauchy--characteristic extraction~(CCE)~\cite{Reisswig:2009us,Mitman:2020pbt,Moxon:2021gbv}, and matching has evolved merging binaries with resolved tails~\cite{Ma:2023qjn,Ma:2024hzq}. However, many Bondi-like formulations have weakly hyperbolic gauge sectors, obstructing standard energy estimates~\cite{Giannakopoulos:2020dih,Giannakopoulos:2021pnh}. For two-way matching, the spacelike construction has a structural advantage: the generalized harmonic interior and the hyperboloidal exterior are both symmetric hyperbolic~\cite{Friedrich:1985afv,Lindblom:2005qh,Nutzi:2025kqc}, so they can exchange characteristic fields across the matching surface. In the spacelike construction, outgoing null rays are not coordinate lines, so their caustics do not exclude matching close to the black holes~\cite{Ma:2024hzq}.

\emph{Regular Einstein evolution across fixed null infinity.}---%
The formulation of the Einstein equations in Ref.~\cite{Nutzi:2025kqc} gives a regular system to evolve
nonlinear perturbations of Minkowski spacetime up to and including $\scri$. The  unknown is an orthonormal frame for the metric (16 components), a metric connection (24) and the rescaled Weyl curvature (10).
We write the  unknown as a
reference field plus our evolved unknown $U$.
The gauge condition sets ten temporal frame and connection components
of $U$ to zero, leaving 40 independent components for $U$.
They satisfy a symmetric hyperbolic system:
\begin{equation}
 A^0(U,x)\partial_\tau U+A^i(U,x)\partial_{\chi^i}U=S(U,x).
 \label{eq:regular-hyperbolic}
\end{equation}
The coordinates $x=(\tau,\chi^i)$ and all coefficients are regular across $\scri$. The matrices $A^\mu$ are real symmetric, with $A^0$ uniformly positive definite for small $U$~\cite{Nutzi:2025kqc}.
The Einstein equations are \eqref{eq:regular-hyperbolic} together with 36 spatial constraints that propagate by the Maurer--Cartan structure.

The equations are formulated on the conformal compactification of Minkowski spacetime, and the location of $\scri$ for
each such perturbation is the location of $\scri$ for Minkowski itself~\cite{Nutzi:2025kqc}.
We can thus choose coordinates in which the cuts of $\scri$
are
at a time-independent grid radius. Both homothetic and stationary coordinates achieve this~\cite{Zenginoglu:2026bwc}.
For extraction and matching we prefer dimensionless stationary hyperboloidal coordinates $(\tau,\rho)$~\cite{Zenginoglu:2007jw}, related to the reference Minkowski coordinates $(t,r)$ by
\begin{equation}
 t=L\tau+\sqrt{L^2+r^2},\qquad
 r=\frac{2L\rho}{1-\rho^2}.
 \label{eq:stationary-compactification}
\end{equation}
Here $L$ is the curvature radius of the reference hyperboloids.
These coordinates keep both the worldtube ($\rho=\rho_\Gamma$) and $\scri$ ($\rho=1$) fixed, unlike homothetic coordinates.
Our formulation and gauge condition are invariant under $\tau$-translations.
For extraction we use a stationary Schwarzschild temporal reference field. For matching we add a correction linear in time, choosing its initial value and derivative to agree with the perturbed interior at the worldtube.
At the boundary $\scri$, characteristics leave the domain or run along
$\scri$; no boundary data are needed.

In our black-hole applications we use \eqref{eq:regular-hyperbolic} only in an exterior hyperboloidal layer~\cite{Zenginoglu:2010cq,Bernuzzi:2011aj,Bernuzzi:2025hhu}. For two-way matching, a generalized harmonic solver evolves the strong-field interior. Our perturbed-black-hole evolution tests the stability of this coupling numerically.

For two-way matching, the interior spatial metric, extrinsic curvature, and gauge determine the frame, connection, Weyl curvature, and their derivatives where the initial slice meets the worldtube.
We extend these fields to $\scri$ by solving the nonlinear constraints on the initial exterior slice (see \hyperref[em:matched-initial-data]{End Matter}). For binary waveform extraction, we instead prescribe Schwarzschild data there.

At $\scri$, we follow the null generators and integrate the Bondi
retarded time $u$ along them using the reconstructed conformal scale.
We evaluate the radiative fields on common constant-$u$ cuts in a
polarization basis transported along the generators%
~\cite{Frauendiener:1998yi,Frauendiener:2021kqm}.
The strain $C=\lim_{r_B\to\infty}r_B(h_++ih_\times)$, with Bondi areal radius $r_B$, follows from the transverse null shear~\cite{Bishop:2013ona}, evaluated from frame derivatives or the evolved connection. The radiation flux is determined by the Bondi news $N=\partial_uC$; the evolved Weyl tensor provides
the spin-$2$ curvature:
\begin{equation}
 q=-\overline{\Psi_4^0}=\partial_uN=\partial_u^2C.
 \label{eq:curvature-waveform}
\end{equation}
We denote spin-weighted spherical-harmonic coefficients by $q_{\ell m}$ and $C_{\ell m}$.

\begin{figure}[t]
  \centering
  \includegraphics[width=\columnwidth]{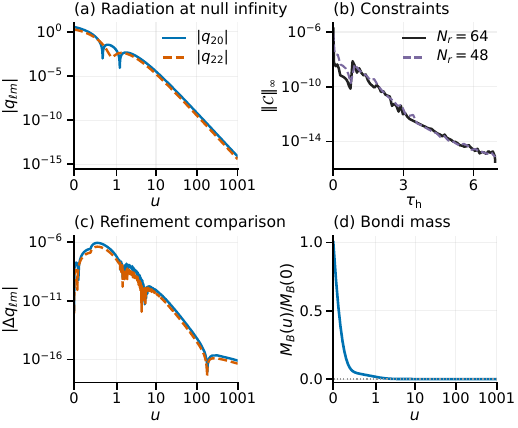}
  \caption{Dispersion of H\"ubner's nonlinear vacuum perturbation ($\varepsilon=0.5$) in homothetic scri-fixed coordinates.
  (a) Bondi curvature modes.
  (b) Absolute maximum of all 36 constraint residuals on the evolution
  grid, versus homothetic time $\tau_{\rm h}$; $N_r$ is the number of radial points.
  (c) Absolute differences between coarse and fine curvature modes on common Bondi cuts under joint spatial and temporal refinement.
  (d) Flux-integrated Bondi mass $M_B$. On the final cut it differs from the
  independently reconstructed mass-aspect estimate by $1.26\times10^{-6}$
  of the initial Bondi mass.
  The time axes in (a,c,d) are logarithmic in $u+0.1$ and labeled by $u$.}
  \label{fig:controlled-evolution}
\end{figure}

\begin{figure*}[tp]
  \centering
  \includegraphics[width=\textwidth]{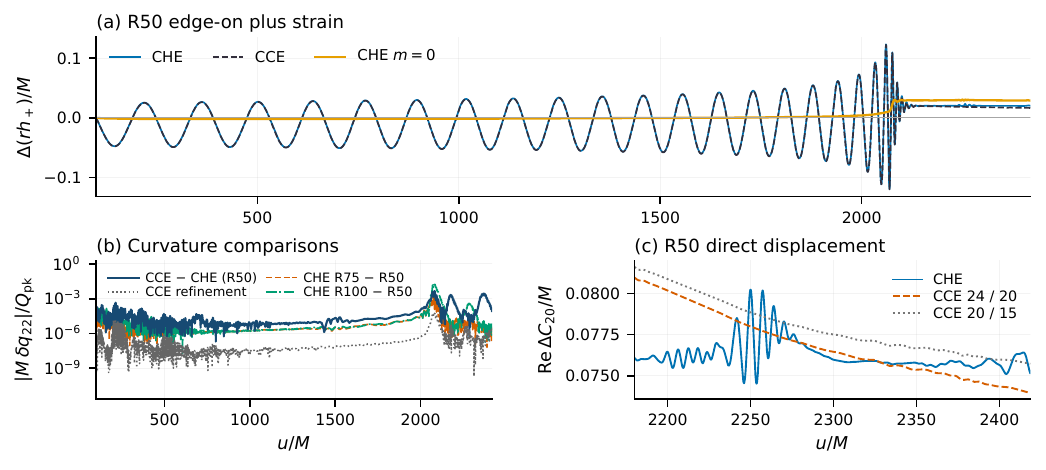}
  \caption{Two extractions of a ten-orbit binary merger at null infinity. (a) Edge-on plus polarization from R50 at $(\theta,\varphi)=(\pi/2,0)$, using the $\ell=2$, $m=0,\pm2$ modes. Orange isolates the CHE $m=0$ contribution. (b) Pointwise differences in the complex curvature mode $q_{22}$, normalized by the R50 CHE peak $Q_{\rm pk}=\max|Mq_{22}^{\rm CHE,R50}|$. The curves compare extraction methods, worldtube radii, and CCE resolutions. (c) Postmerger displacement $\operatorname{Re}\Delta C_{20}/M$ at R50. CCE labels give the evolution angular cutoff and radial point count. All strain changes are measured from $u=100M$.}
  \label{fig:bbh-extraction}
\end{figure*}

\emph{Radiative perturbations of Minkowski spacetime.}---%
We use H\"ubner's nonsymmetric conformal metric~\cite{Hubner:2000zn,Hubner:2000pb} as free data. We construct initial data separately for amplitudes $\varepsilon=+0.5$ and $-0.5$ by solving the degenerate elliptic Yamabe equation of~\cite{Andersson:1992yk}.
We use Dedalus~\cite{Burns2020Dedalus} to evolve Eq.~\eqref{eq:regular-hyperbolic} on a ball in homothetic scri-fixed coordinates. To isolate the part even in $\varepsilon$, we average the two strain displacements.

The curvature amplitudes $\lvert q_{20}\rvert$ and $\lvert q_{22}\rvert$ at null infinity fall from order unity to $\sim10^{-14}$ [Fig.~\ref{fig:controlled-evolution}(a)]. For either sign and at both resolutions, fits over $400\leq u\leq800$ give exponents near $-4.98$, consistent with the $u^{-5}$ decay expected from regularity at future timelike infinity~\cite{Friedrich:1986qfi}. The constraint residuals fall to near machine precision [Fig.~\ref{fig:controlled-evolution}(b)]. Refinement changes each complex mode by less than $6\times10^{-7}$ in relative $L^2$ norm over the full evolution. Even over
$900\leq u\leq1001$, the two resolutions agree to within $1\%$ in this norm, with absolute differences below $10^{-16}$ on every sampled cut [Fig.~\ref{fig:controlled-evolution}(c)].

At H\"ubner's amplitude ($\varepsilon=1$), our $M_B(u)/M_B(0)$ agrees with his digitized values to within $7\times10^{-3}$ without alignment through $u=2$, when his solution has radiated over $99\%$ of its initial Bondi mass~\cite{Hubner:2000pb}. H\"ubner left open whether longer integration would remove the late-time mass offset or whether it reflected an error in the initial Bondi mass. As his cuts of $\scri$ approached future timelike infinity $i^+$, fewer grid points remained in the physical domain.
We keep $\scri$ at fixed radius and leave $i^+$ at infinite evolution time, retaining the full radial grid as the waves disperse (for an alternative treatment of $i^+$, see Ref.~\cite{Koukas:2026gup}). At $u=1001$ we measure the remaining mass directly from the mass aspect, independently of the flux integral in Fig.~\ref{fig:controlled-evolution}(d). This resolves H\"ubner's late-time question for our $\varepsilon=\pm0.5$ evolutions: $|M_B|/M_B(0)<2\times10^{-10}$ for both signs, consistent with complete dispersal.

The radiation produces an accumulated displacement between early
and late Bondi cuts. Its electric-parity part $\Delta C_E$ obeys the vacuum
Bondi balance law~\cite{Mitman:2020pbt},
\begin{equation}
 \Delta C_E=\mathcal K\!\left[
     \int_{u_0}^{u_1}|N|^2\,du+4\Delta m_B\right].
 \label{eq:memory-balance}
\end{equation}
The news-squared term describes nonlinear radiation-flux
memory~\cite{Christodoulou:1991cr}; the change in the Bondi mass aspect $m_B$
contributes to the Coulombic displacement. The inverse angular operator
$\mathcal K$ maps these sources to spin-two strain.
From $u=49$ to the final cut, the even-in-$\varepsilon$ electric-parity displacement changes by only $0.021\%$ in angular $L^2$ norm over $2\leq\ell\leq8$. The radiation flux alone would overestimate the final displacement norm by a factor of $3.5$. The change in mass aspect largely cancels the flux contribution, and their sum reproduces the displacement measured from the metric [Fig.~\ref{fig:em-huebner-memory}]. The pulse thus leaves a lasting nonlinear memory at null infinity even when the remaining Bondi mass is indistinguishable from zero.

\begin{figure*}[tp]
  \centering
  \includegraphics[width=\textwidth]{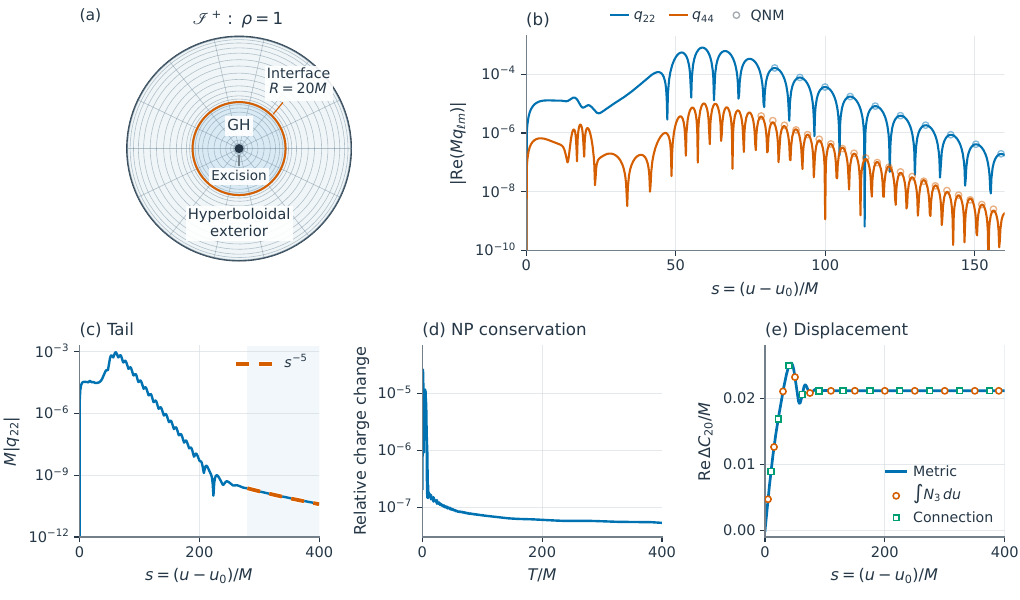}
  \caption{
  Cauchy--hyperboloidal matching for a rotating quadrupolar perturbation of Schwarzschild with initial amplitude $A=0.01$.
  (a) Computational domain with excision sphere at $1.6M$ and matching worldtube at $20M$.
  The figure uses $\rho$ [Eq.~\eqref{eq:stationary-compactification}],
  with $L=20M$, also in the interior.
  (b) The peaks of $\lvert\operatorname{Re}(Mq_{\ell m})\rvert$ follow the spacing and decay predicted for the fundamental Schwarzschild quasinormal modes (circles)~\cite{Leaver:1985ax}.
  (c) Raw $|Mq_{22}|$ and an $s^{-5}$ guide for a nonzero quadrupolar charge.
  (d) The relative change $\|G(T)-G(0)\|_2/\|G(0)\|_2$ of the Newman--Penrose charge vector $G$
  peaks at $2.66\times10^{-5}$ during startup ($T=1.2M$) and remains below $1.9\times10^{-7}$ for $10M\leq T\leq400M$.
  (e) Predominantly Coulombic displacement $\operatorname{Re}\Delta C_{20}/M$ between cuts. We compute the transverse null shear using the connection reconstructed from frame derivatives (line) or the evolved connection (squares). The circles integrate the news $N_3$, reconstructed from $\Psi_3^0$ by angular inversion of the Bianchi identity.
  Panels (b,c,e) use Bondi time $s=(u-u_0)/M$; (d) uses
  evolution time $T=L(\tau-\tau_0)$.}
  \label{fig:matching}
\end{figure*}

\emph{Cauchy-hyperboloidal extraction.}---The phase of a binary-black-hole waveform helps astronomers determine the black
holes' masses and spins~\cite{Cutler:1994ys,LIGOScientific:2016vlm}.
The radiation can also leave a memory that persists after the
oscillations fade~\cite{Christodoulou:1991cr,Mitman:2020pbt}.
Finite-domain binary simulations record these waves on spheres, each tracing a timelike worldtube.
In Cauchy--hyperboloidal extraction (CHE), worldtube data drive a nonlinear exterior evolution on hyperboloidal slices reaching $\scri$, without feedback to the interior. This lets us compare CHE and CCE on the same data from an existing simulation.

We demonstrate CHE on three worldtubes ($R=50M,75M,100M$, labeled R50, R75, R100) from a ten-orbit, equal-mass, initially nonspinning binary evolved with \GRA{}~\cite{Daszuta:2021ecf} [Fig.~\ref{fig:bbh-extraction}]. The time-dependent worldtube geometry and Weyl tensor provide the boundary data for our evolution equation~\eqref{eq:regular-hyperbolic}. For the evolution, we developed a GPU pseudospectral solver on a spherical shell with global angular harmonics and fourth-order exponential time-differencing Runge--Kutta integration~\cite{CoxMatthews2002}.

The CHE curvature waveform is a projection of the evolved Weyl tensor. Across the three worldtubes, the dominant complex mode $q_{22}$ differs from refined SpECTRE CCE~\cite{Moxon:2021gbv} by $0.22\%$--$0.54\%$ in relative $L^2$ norm over the merger interval $1900\leq u/M\leq2200$, without fitting a time shift, phase, or amplitude [Fig.~\ref{fig:bbh-extraction}(b)].

Strain provides a more demanding comparison: its reconstruction requires Bondi cuts, a conformal scale, and a polarization frame at $\scri$~\cite{Hubner:2000pb,Bishop:2013ona,Handmer:2016mls}. In CHE we obtain the strain from the transverse null shear in that frame, computed from the evolved connection. Over the merger interval at R50, the complex strain changes $\Delta C_{22}$ from CHE and refined CCE differ by $4.6\%$ in relative $L^2$ norm, but the edge-on plus polarizations differ by only $1.3\%$: the $m=0$ and $m=\pm2$ differences partly cancel in this projection [Fig.~\ref{fig:bbh-extraction}(a)]. On the final cut, most of the edge-on strain difference comes from $m=\pm2$, whereas the $(2,0)$ displacements are similar despite different postmerger drifts [Fig.~\ref{fig:bbh-extraction}(c)].

At R50, the maximum CHE--CCE difference in the $(2,0)$ radiation-flux contribution to memory [Eq.~\eqref{eq:memory-balance}] over the common cuts is $0.031\%$ of the magnitude of its final refined CCE value. CHE and CCE thus agree on both the dominant merger curvature and the nonlinear radiation-flux memory. On the final R50 cut, CHE satisfies the integrated Bondi mass-loss law to within $0.37\%$ of the radiated energy. The refined CCE output instead gives a mass increase and cannot independently check the Coulombic contribution (see \hyperref[em:binary-budget]{End Matter}).

For CHE we prescribe Schwarzschild data on the initial hyperboloidal slice, estimating the mass from the monopole on the outermost sphere in the Cauchy data. These data satisfy the vacuum constraints but contain no inspiral radiation and do not match the nonspherical binary geometry where the slice meets the worldtube. This mismatch can contaminate the early waveform. Future work should construct exterior data compatible with the binary worldtube (see Ref.~\cite{DaRe:2026wyd} for related work in CCE).

\emph{Cauchy--hyperboloidal matching.}---%
CHE leaves the Cauchy evolution's artificial outer boundary in place, where it can reflect radiation. Cauchy--hyperboloidal matching (CHM) makes the worldtube an internal interface instead. We test this construction for a Schwarzschild black hole of mass $M$ perturbed by a rotating quadrupolar wave packet.

Inside the worldtube we evolve the full nonlinear equations in generalized harmonic (GH) gauge~\cite{Friedrich:1985afv}, using the first-order system of Ref.~\cite{Lindblom:2005qh} with excision inside the horizon. Outside the worldtube we evolve Eq.~\eqref{eq:regular-hyperbolic} in stationary hyperboloidal coordinates out to $\scri$ [Fig.~\ref{fig:matching}(a)].

At each Runge--Kutta stage, coordinate, conformal, and Lorentz transformations relate the interface fields, as in Cauchy--characteristic matching~\cite{Ma:2023qjn}. Each side supplies curvature data for the other's two incoming radiative degrees of freedom. With the exterior reference field prescribed, the interior metric and its first derivatives fix the incoming exterior frame and connection fields. Local differential conditions control the six remaining incoming exterior Weyl components and the incoming GH constraints; a separate condition fixes the incoming GH gauge modes.

A nonlinear constraint solver extends the interior initial data to $\scri$ with compatible curvature and derivatives at the interface. The initial exterior perturbation is noncompact. Our GPU matching code advances two pseudospectral shells with global angular harmonics using classical fourth-order Runge--Kutta. We damp the GH constraints in the interior; angular filtering preserves complete tensor multipoles on both shells. At $T=400M$, the root-mean-square residuals are $2.9\times10^{-8}$ for the interior harmonic constraints, $5.1\times10^{-8}$ for the exterior Weyl constraints, and $1.2\times10^{-7}$ for the remaining 30 exterior constraints combined.

At $\scri$, the dominant $(2,2)$ and weaker $(4,4)$ curvature modes ring at frequencies consistent with Schwarzschild quasinormal modes~\cite{Leaver:1985ax} [Fig.~\ref{fig:matching}(b)]. A power-law tail follows~\cite{Price:1971fb,DeAmicis:2024eoy}. For linearized gravity on Schwarzschild, $q_{22}$ decays as $u^{-6}$ for generic compactly supported data, but as $u^{-5}$ for nonzero quadrupolar Newman--Penrose (NP) charge~\cite{Ma:2021cno}. We measure the five complex NP charges independently of the outgoing waveform, from the subleading quadrupolar Weyl curvature at $\scri$~\cite{Newman:1968uj,Dain:2001bh,Frauendiener:2023ltp} (see \hyperref[eq:np-charge]{End Matter}). The $m=\pm2$ charges are nonzero, and the charge vector remains constant to numerical accuracy after startup [Fig.~\ref{fig:matching}(d)]. The nonlinear tail is consistent with the nonzero-charge prediction: the fitted slope of $|q_{22}|$ is $-4.86$ over $280\leq (u-u_0)/M\leq400$, with $u_0$ fixed at the initial Bondi cut [Fig.~\ref{fig:matching}(c)].

Three reconstructions agree on $\operatorname{Re}\Delta C_{20}\simeq0.0212M$ between the initial and final cuts [Fig.~\ref{fig:matching}(e)]. The Bondi balance [Eq.~\eqref{eq:memory-balance}] attributes almost all of this displacement to the change in mass aspect; the radiation flux contributes only $7.15\times10^{-6}M$.

\Needspace{4\baselineskip}
\emph{Discussion.}---%
Computing astrophysical waveforms by evolving the nonlinear Einstein equations on spacelike slices reaching future null infinity has been a goal of numerical relativity since the 1990s. We achieve this goal with regular equations, scri-fixing, and hyperboloidal layers. For a dispersing vacuum perturbation of Minkowski spacetime, we measure both the Bondi mass loss and the nonlinear memory left by the radiation. For a ten-orbit binary black hole merger, we extract the waveform, whose dominant curvature mode agrees with characteristic extraction to better than one percent through merger. For a perturbed black hole, two-way matching connects the strong-field interior to $\scri$ in a single evolution, resolving quasinormal ringing and the curvature tail.

Hyperboloidal compactification with scri-fixing~\cite{Zenginoglu:2007jw} is widely used in perturbation theory on fixed black-hole backgrounds~\cite{Bernuzzi:2025hhu,DeAmicis:2024eoy,Besson:2024adi,Bourg:2025lpd,Vaswani:2026twx,PanossoMacedo:2024nkw,Baptista:2025kmo,siahmazgi2026}. It tracks radiation at null infinity from the initial burst to the late-time tail. Our work extends this approach to the full nonlinear Einstein equations, where the geometry carrying the waves is itself dynamical. This opens the prospect of making hyperboloidal methods as routine in numerical relativity as they are in black-hole perturbation theory.

The next step is to extend the two-way matching demonstrated here for a single perturbed black hole to binary inspiral and merger. If successful, the Cauchy grid could end near the binary, with a hyperboloidal layer of finite coordinate thickness covering the entire exterior out to $\scri$. Replacing the extensive grid between the binary and a distant artificial boundary could reduce the cost of long inspirals. More fundamentally, it would change how numerical relativity predicts gravitational waves: the waveform and its memory at null infinity would be direct outputs of the same evolution that follows the black holes through merger.

\emph{Acknowledgments.}---A.Z. is supported by the National Science Foundation under Grant
No.~2309084.
S.B. acknowledges fundings from the EU Horizon under ERC Consolidator Grant no. InspiReM-101043372
and from the Deutsche Forschungsgemeinschaft (DFG) project ``GROOVHY''
(BE 6301/5-1 Projektnummer: 523180871).

A.Z. thanks the Department of Mathematics at the University of Maryland for access to its high-performance computing resources.
Computations were also performed on the ARA cluster at Friedrich Schiller University Jena and on the \texttt{Tullio} cluster at INFN Torino.
The authors gratefully acknowledge the Gauss Centre for Supercomputing
e.V. (\url{www.gauss-centre.eu}) for funding this project by providing
computing time on the GCS Supercomputer SuperMUC-NG Phase 2 at Leibniz
Supercomputing Centre (\url{www.lrz.de}) through project \texttt{pn67xo}.
The ARA cluster is funded in part by DFG grants INST
275/334-1 FUGG and INST 275/363-1 FUGG, and ERC Starting Grant, grant
agreement no. BinGraSp-714626.

\emph{Data Availability.}---We plan to deposit the numerical data underlying the figures in a public repository upon publication. We used Dedalus~\cite{Burns2020Dedalus} for the Minkowski benchmark, \GRA{}~\cite{Daszuta:2021ecf,Cook:2023bag} for the binary evolution, and SpECTRE~\cite{Moxon:2021gbv,spectrecode} for characteristic extraction.

\emph{AI Use Disclosure.}---OpenAI Codex assisted with code development, analysis of the results, figure preparation, and revising the manuscript. The authors take full responsibility for the results and their interpretation.

\bibliography{references}

\clearpage
\section*{End Matter}

\phantomsection\label{em:regular-equations}%
\emph{Evolution.}---In the formulation of Ref.~\cite{Nutzi:2025kqc}, the Einstein equations take the form of a Maurer--Cartan equation in a differential graded Lie algebra:
\begin{equation}
 \mathcal R=\mathsf{d}\,\mathcal U+\tfrac12[\mathcal U,\mathcal U]=0.
 \label{eq:maurer-cartan}
\end{equation}
The differential $\mathsf d$ describes linearized gravity about Minkowski spacetime; the bracket describes the nonlinear interaction.
The unknown $\mathcal U$ has two parts: a Poincar\'e-algebra-valued one-form describing the frame and connection, and the rescaled Weyl tensor $d^a{}_{bcd}$.
The solution $\mathcal{U}=0$ of Eq.~\eqref{eq:maurer-cartan} is Minkowski spacetime.

We write $\mathcal U=\mathcal U_0+U$.
The prescribed reference field $\mathcal U_0$ vanishes for the Minkowski benchmark; for extraction it is given by the Schwarzschild temporal frame and connection components; for matching we add a correction fixed by the interior data and their first time derivative
on the initial worldtube cut.
The gauge condition requires that the frame and connection one-form of $U$ vanishes on the timelike vector field
$\partial_\tau+\rho\partial_\rho$,
which points outward at $\scri$.
These ten gauge conditions
are independent of $\mathcal{U}_0$ and
leave 40 independent components for
$U$. Of the 76 equations in Eq.~\eqref{eq:maurer-cartan}, 40 give the evolution system~\eqref{eq:regular-hyperbolic}; the remaining 36 give the spatial constraints after eliminating time derivatives. Constraint propagation follows from the identity $\mathsf d\mathcal R+[\mathcal U,\mathcal R]=0$, which holds for every field $\mathcal U$. For the Minkowski linearization, the Weyl-constraint energy decays with homogeneous incoming constraint data.

\phantomsection\label{em:matched-initial-data}%
\emph{Hyperboloidal initial data for matching.}---We seek $U_\Sigma=U|_{\tau=0}$ on the exterior initial slice. The interior determines the geometry on the matching sphere $\Gamma$, but not a unique exterior. The spatial metric $\gamma_{ij}$ and extrinsic curvature $K_{ij}=-\tfrac12\mathcal L_N\gamma_{ij}$, with future unit normal $N$, give the electric and magnetic Weyl tensors in vacuum:
\begin{equation}
 \begin{aligned}
 E_{ij}&=R_{ij}[\gamma]+\kappa K_{ij}-K_i{}^kK_{kj},\\
 B_{ij}&=\epsilon_i{}^{kl}D_kK_{lj},\qquad
 \kappa=\gamma^{ij}K_{ij}.
 \end{aligned}
 \label{eq:initial-curvature}
\end{equation}
Changing the normal mixes $E$ and $B$. The curvature in the exterior tetrad, with its conformal weight, and the transformed frame and connection coefficients fix $U_\Gamma=U_\Sigma|_\Gamma$. The interior gauge and vacuum evolution supply the time derivatives needed to transform derivatives across $\Gamma$. For related constructions see Refs.~\cite{Buchman:2009ew,Schinkel:2013zm,Csukas:2026wpa}.

The extension must satisfy the 36 spatial constraints $\mathcal C_\Sigma(U_\Sigma)=0$. Let $U_0$ consist of the spatial frame and connection components and all ten Weyl components of Schwarzschild. At $U_0$ the radial derivative coefficient $P_0^\rho$ has rank 26 and nullity 14. Write
\begin{equation}
 U_\Sigma=U_0|_{\tau=0}+D\zeta+Ff,\qquad P_0^\rho F=0,
 \label{eq:initial-split}
\end{equation}
where $F$ spans the kernel and $D$ a complement. The 14 free fields $f$ (ten geometric and four transverse trace-free curvature components) have decaying Hermite profiles matching their interior values and derivatives at $\Gamma$.
For the 26 fields $\zeta$ we solve
\begin{equation}
 \Pi_0\mathcal C_\Sigma(U_\Sigma)=0,\qquad U_\Sigma|_\Gamma=U_\Gamma,
 \label{eq:initial-range}
\end{equation}
where $\Pi_0$ projects onto the rank-26 range of $P_0^\rho$.
In a refined initial-data solve,
the coefficient of $\partial_\rho\zeta$ in Eq.~\eqref{eq:initial-range}
was invertible
at every sampled point.
Each iteration solves a linear inhomogeneous problem,
where the linear operator is the Schwarzschild-linearization
of Eq.~\eqref{eq:initial-range} with respect to $\zeta$,
using spectral collocation over the whole shell.
At $\scri$ we require regularity without prescribing boundary data. We solve for curvature directly; the whole-ball construction below recovers it from the initial metric.

Smooth data require corner compatibility: at $\Gamma$ both descriptions must give the same field and normal derivatives in the same coordinates and frame, with the same conformal normalization. Only the free derivatives are prescribed. Differentiating the constraints determines the others successively; we check them against the transformed interior derivatives.

On a smooth solution of Eq.~\eqref{eq:initial-range}, the spatial Bianchi identity transports the ten remaining residuals $c$ homogeneously along $v$,
the radial leg of the exterior frame projected onto the slice along its timelike leg. Thus, if $v^\rho>0$ throughout the closed shell, $c|_\Gamma=0$ implies $c=0$ up to $\scri$. We found $v^\rho>0$ at every sampled point. For the smooth quadrupolar perturbation of Schwarzschild evolved here, all 36 residuals were below $4.6\times10^{-12}$ at $\Gamma$ and $4.0\times10^{-10}$ on an independent shell grid. The radial equations for $\zeta$ can amplify high angular frequencies; transport of $c$ does not establish their well-posedness for arbitrary data.

\emph{Initial data for radiative perturbations of Minkowski spacetime.}---H\"ubner's initial slice fills the compact unit ball $r_{\rm H}^2=x^2+y^2+z^2\leq1$, with $\scri$ at its boundary and no inner matching sphere. The free data are a conformal metric on the whole ball~\cite{Hubner:2000pb}:
\begin{equation}
 \begin{aligned}
 h&=\left[1+\tfrac{\varepsilon}{3}b^2(x^2+2y^2)\right]dx^2+dy^2+dz^2,\\
 b&=\tfrac{1}{2}(1-r_{\rm H}^2).
 \end{aligned}
 \label{eq:huebner-seed}
\end{equation}
Its umbilic boundary satisfies the
smoothness conditions of~\cite{Andersson:1992yk}. With constant pure-trace extrinsic curvature, the momentum constraint holds identically. The Hamiltonian constraint gives a nonlinear Yamabe equation for $\phi$, which determines the initial conformal factor $\Omega_{\rm H}=b/\phi^2$ and the physical metric $\widetilde h=\Omega_{\rm H}^{-2}h$. H\"ubner's sign $\widetilde K^{\rm H}_{ij}=+\tfrac12\mathcal L_N\widetilde h_{ij}$ gives trace $+3$, hence $\kappa=-3$ in Eq.~\eqref{eq:initial-curvature}. The initial magnetic Weyl field vanishes. We reconstruct the rescaled electric Weyl field from the Ricci tensor of $h$ and the Hessian of $\Omega_{\rm H}$, factoring out the known boundary zeros before evaluating the formally singular quotients.

We solve the Yamabe equation separately for $\varepsilon=\pm0.5$
on a grid with $(N_r,N_\theta,N_\varphi)=(80,40,80)$. For each sign we evolve the data on two grids, $(48,24,24)$ and $(64,32,64)$, with fine-to-coarse timestep ratio $9/16$. We compare the complex curvature modes on common cuts, normalizing each difference by the time-$L^2$ norm of the finer solution for that mode, without alignment or extrapolation. The mass aspect gives $|M_B|<2\times10^{-10}M_B(0)$ at $u=1001$, independently of the flux history in Fig.~\ref{fig:controlled-evolution}(d).

\emph{Measuring radiation and displacement.}---We measure strain $C$ and outgoing curvature $q$ on the same Bondi cuts in a common polarization frame. Let $\bar g$ be the regular conformal metric determined by the frame. The defining function $\Omega$ of $\scri$ gives its null generator $\boldsymbol{k}=\bar g^{-1}d\Omega$. On each cut, let $\boldsymbol m$ be the complex tangent dyad and $\boldsymbol\ell$ the transverse null normal, with $\bar g(\boldsymbol m,\overline{\boldsymbol m})=1$ and $\bar g(\boldsymbol{k},\boldsymbol\ell)=-1$. The conformal scale $\mathcal A$ fixes the Bondi metric $g_B=\mathcal A^{-2}\bar g$ and clock $(\mathcal A\boldsymbol{k})(u)=L$. In Eq.~\eqref{eq:stationary-compactification}, $\Omega=(1-\rho^2)/2$ and $L$ is the curvature radius; for H\"ubner, $\Omega=e^{-\tau_{\rm h}}(1-\rho^2)$ and $L=2$ is the length unit.
The shear of $\boldsymbol\ell$ gives the strain; a projection of the evolved Weyl tensor gives the curvature waveform:
\begin{equation}
 C=-2L\mathcal A\,\boldsymbol m^a\boldsymbol m^b\bar\nabla_a\boldsymbol\ell_b,\;\;
 q=-\tfrac{2\mathcal A^3}{L}d(\boldsymbol{k},\boldsymbol m,\boldsymbol{k},\boldsymbol m).
 \label{eq:direct-strain}
\end{equation}
Here $\bar\nabla$ and index lowering refer to $\bar g$. In our Bondi tetrad $q=-\overline{\Psi_4^0}$, where $\Psi_4^0=\lim_{r_B\to\infty}r_B\Psi_4$, so $q=\partial_u^2 C$ [Eq.~\eqref{eq:curvature-waveform}]. We project both fields onto spin-two harmonics. For the Minkowski benchmark we reconstruct the connection from frame derivatives; for binary extraction we use the evolved connection.

\phantomsection\label{em:memory-balance}%
Two sources enter the displacement balance [Eq.~\eqref{eq:memory-balance}]: radiated energy and the change in mass aspect $m_B=-\operatorname{Re}(\overline{\Psi_2^0}+N\bar C/4)$. The scalar-to-spin-two weight of $\mathcal K$ is $[(\ell-1)\ell(\ell+1)(\ell+2)]^{-1/2}$ for $\ell\geq2$. Because $m_B$ contains the absolute strain $C$, balance tests the measured geometry for consistency.

\begin{figure}[!ht]
  \centering
  \includegraphics[width=\columnwidth]{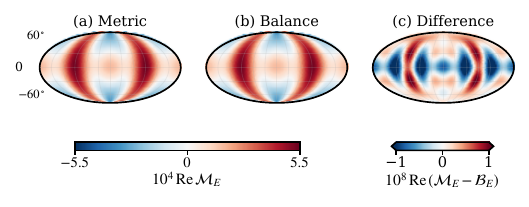}
  \caption{H\"ubner memory on the Bondi sphere at $u=1001$ ($2\leq\ell\leq8$). The metric displacement $\mathcal M_E=[\Delta C_E(+\varepsilon)+\Delta C_E(-\varepsilon)]/(2\varepsilon^2)$ (left) and the same combination $\mathcal B_E$ from the balance law (center) are scaled by $10^4$. Their difference (right) is scaled by $10^8$.}
  \label{fig:em-huebner-memory}
\end{figure}

\phantomsection\label{em:huebner-checks}%
For the radiative Minkowski benchmark, refinement tests the even-in-$\varepsilon$ metric displacement itself: its relative angular $L^2$ change is $9.89\times10^{-5}$ through $\ell=8$ and $8.83\times10^{-4}$ through $\ell=11$. At $u=1001$ its relative angular $L^2$ difference from the balance prediction over $2\leq\ell\leq8$ falls from $1.24\times10^{-4}$ to $2.23\times10^{-5}$ [Fig.~\ref{fig:em-huebner-memory}].

\phantomsection\label{em:binary-budget}%
\emph{Flux memory and total displacement.}---CHE and CCE agree better on the binary's radiation-flux memory than on its total displacement. We form $|N|^2$ from all news modes through $\ell=8$ and project onto $(2,0)$. From $u=100M$ to the last common cut at $u\simeq2419M$, the maximum difference between the CHE and refined CCE flux contributions is $0.031\%$ at R50 and $0.047\%$ at R75. At R100 it is more than an order of magnitude larger, $0.76\%$. Each comparison is normalized by the final refined CCE flux contribution on the same worldtube.

\begin{table}[!h]
\caption{R50 binary displacement and contributions to its balance for $(\ell,m)=(2,0)$, in units of $M$, from $u=100M$ to $u\simeq2419M$. CHE uses 128 radial points, $J_{\max}=32$, and $\Delta\tau=0.002$. CCE refinement increases the radial points from 15 to 20, the input angular band from 16 to 20, and the evolution band from 20 to 24.}
\label{tab:em-binary-budget}
\begin{ruledtabular}
\begin{tabular}{lrrr}
 & Displacement & Radiation flux & Coulombic \\
CHE         & $0.07516$ & $0.08164$ & $-0.00641$ \\
Baseline CCE & $0.07572$ & $0.08165$ & $-0.00167$ \\
Refined CCE & $0.07395$ & $0.08165$ & $-0.31041$ \\
\end{tabular}
\end{ruledtabular}
\end{table}

At R50, CHE and both CCE resolutions give similar displacements and flux contributions (Table~\ref{tab:em-binary-budget}). Refining CCE barely changes the flux, but the Coulombic term grows to several times the measured displacement in magnitude. Both methods give $E_{\rm rad}\simeq0.03986M$; the final residual $\Delta M_B+E_{\rm rad}$ is $1.48\times10^{-4}M$ for CHE and $0.17135M$ for refined CCE. Recomputing the mass aspect from the saved CCE modes traces the apparent mass increase to the monopole of $\Psi_2^0$ and the large Coulombic displacement to its $(2,0)$ mode. This output cannot independently validate the Coulombic contribution.

\emph{Newman--Penrose charges and late-time tail.}---As radiation crosses $\scri$, the Bondi mass decreases. The
Newman--Penrose charges
remain constant and contain information about the asymptotic curvature of the
initial data.

In a canonical Bondi tetrad, the charges project out the quadrupole of the $r_B^{-6}$ coefficient of the incoming Weyl scalar $\widetilde\Psi_0$. We evaluate them directly on a cut $\mathcal S$ of $\scri$ from the rescaled Weyl tensor $d$ and the cut geometry~\cite{Frauendiener:2023ltp}:
\begin{equation}
 \begin{aligned}
 G_j&=-\int_{\mathcal S}\!\mathcal V_j\mathcal F\,dS,
       \qquad j=-2,\ldots,2,\\
 \mathcal F&=(\bar\nabla_{\boldsymbol\ell}d)(\boldsymbol\ell,\boldsymbol m,\boldsymbol\ell,\boldsymbol m)
              +5\theta_{\boldsymbol\ell}\,d(\boldsymbol\ell,\boldsymbol m,\boldsymbol\ell,\boldsymbol m).
 \end{aligned}
 \label{eq:np-charge}
\end{equation}
Here $\theta_{\boldsymbol\ell}=\overline{\boldsymbol m}^{\,a}\boldsymbol m^b\bar\nabla_a\boldsymbol\ell_b$ is half the expansion along the transverse null normal $\boldsymbol\ell$. The angular weights $\mathcal V_j$ reduce to spin $-2$ quadrupolar harmonics in a round Bondi frame. We solve for these weights on the initial cut and transport them along the generators without renormalization, comparing charges in the same basis throughout.

On the initial cut ($T=0$), $G_2/M^4\simeq57.63+139.56i$ and $G_{-2}\simeq\overline{G_2}$.
In the normalization of Fig.~\ref{fig:matching}(d), halving the timestep changes the charge vector by at most $1.17\times10^{-9}$ through $T=2M$; doubling the time between diagnostic cuts changes it by at most $3.98\times10^{-8}$ through $400M$.

\end{document}